\documentclass[aps,prc,twocolumn,amsmath,amssymb,nofootinbib]{revtex4-2}
\usepackage{graphicx,epsfig}% Include figure files
\usepackage{bm}% bold math
\usepackage{braket}
\usepackage{amsmath}	% required for `\align' (yatex added)
\usepackage{ulem}
\usepackage{color}   %May be necessary if you want to color links
\usepackage[bookmarks=true]{hyperref}
\hypersetup{
	colorlinks=true,
	pdfborder={0 0 0},
	citecolor=blue,
	linkcolor=blue,
	urlcolor=blue
}
\makeatletter
\newcommand{\colorcaption}[2][]{%
	\begingroup%
	\renewcommand{\@caption@fignum@sep}{ (Color online). }%
	\caption[#1]{#2}%
	\endgroup%
}
\makeatother
\newcommand{\nn}{\nonumber}

\begin{document}
	
\title{Enormous nuclear surface diffuseness
  in the exotic Ne and Mg isotopes}
\author{V. Choudhary$^{1}$}
\email{vchoudhary@ph.iitr.ac.in}
\author{W. Horiuchi$^{2}$}
\email{whoriuchi@nucl.sci.hokudai.ac.jp}
\author{M. Kimura$^{2,3,4}$}
\email{masaaki@nucl.sci.hokudai.ac.jp}
\author{R. Chatterjee$^{1}$}
\email{rchatterjee@ph.iitr.ac.in}

\affiliation{$^{1}$Department of Physics, Indian Institute of Technology Roorkee, Roorkee 247 667, India}

\affiliation{$^2$Department of Physics, Hokkaido University, 060-0810 Sapporo, Japan}

\affiliation{$^3$Nuclear Reaction Data Centre, Faculty of Science, Hokkaido University, 060-0810 Sapporo, Japan}

\affiliation{$^4$RIKEN Nishina Center, Wako, Saitama 351-0198, Japan}

\begin{abstract}
\begin{description}
\item[Background] 
The density profile of exotic nuclei can be a rich source of information on the nuclear surface. In particular, the nuclear surface diffuseness parameter is correlated with the occupation probability of nucleons in distinct nuclear orbits, especially those with low angular momenta.

\item[Purpose] The aim of this paper is to investigate the relationship between the nuclear surface diffuseness and spectroscopic information of neutron rich Ne and Mg isotopes both at the cusp and inside the island of inversion.

\item[Method] We use the microscopic antisymmetrized molecular dynamics model to calculate the densities and other spectroscopic information related to Ne and Mg isotopes. A two-parameter Fermi density distribution is then used to define the diffuseness parameter and the matter radius. These quantities are extracted by minimizing the difference between these two densities. To relate them with observables, the two densities are given as inputs to a Glauber model calculation of nucleon-nucleus elastic scattering differential cross section, with the demand that they reproduce the first peak position and its magnitude.

 \item[Results] A marked increase in the occupation of neutrons in the $pf$-orbit is noted in Ne and Mg isotopes from $N=19$ onwards. 
 We observed that the nuclear diffuseness is strongly correlated with the nuclear deformation, in the island of inversion, and gradually increases with the occupation of neutrons in the $1p_{3/2}$ orbit. This result is also confirmed by a single-particle estimate of the valence neutron density distribution, with $^{29}$Ne as a test case. An exception is noted for $^{35-37}$Mg, where the filling up of the holes in the $sd$-shell partially compensates the increase in diffuseness due to filling up of the $1p_{3/2}$ orbit.
 
 \item[Conclusion] Information on nuclear density profile of neutron rich medium mass nuclei can be reliably extracted by studying the first diffraction peak of the nucleon-nucleus elastic scattering differential cross section. The enormous surface diffuseness of Ne and Mg isotopes, in the island of inversion, could be attributed to the increasing neutron occupation of the $1p_{3/2}$ orbit. 
\end{description}	
\end{abstract}

\maketitle

\section{Introduction}

In recent times, investigations of the nuclear surface have revealed
marked distinctions between exotic nuclei and their stable counterparts.
For example, halos \cite{Tanihata85,Tanihata13,Bagchi20}
and skins \cite{Suzuki95} have been seen in nuclei
far from the valley of stability.
The density profile of such exotic nuclei have been an abundant source
of  information on the nuclear surface.
Recently, indication of the core swelling phenomenon
was observed for neutron-rich Ca isotopes,
suggesting that it crucially affects
the density profile near the nuclear surface~\cite{Tanaka20,Horiuchi20}.
Another exotic phenomenon,
a central depression of nuclear density profile,
was reported~\cite{Grasso09,Li16}.
The vacancy in the $s$-orbit plays an essential role
in bubble formation, resulting in a depletion of the central part
of the nuclear density profile. Recently, the present authors
reported that nuclei with bubble-like structure could
have small nuclear surface diffuseness~\cite{Vishal20}.
The nuclear diffuseness
is closely related to the occupation probability near the Fermi level
and increases when the nucleons occupy the low orbital angular momentum state.
This suggests that nuclear surface diffuseness is very sensitive
to the occupation of the nucleons in the distinct nuclear orbits
and therefore a systematic investigation of the nuclear surface
diffuseness is worth pursuing. 

It has been known in the literature that in the medium mass region
there exists a so called ``island of inversion''~\cite{Warburton90},
where intruder configurations with particle-hole excitations
across $N=20$ shell gap in their ground state
result in large deformation. Consequently, deviations from standard shell model
estimates are expected in this region. 
The exotic structure is strongly correlated
with shell evolution and deformation in the island of inversion. 
The nuclear deformations of Ne and Mg isotopes using
the fully microscopic antisymmetrized molecular dynamics (AMD)
with the Gogny-D1S interaction have been analyzed
in Refs.~\cite{Minomo11,Minomo12,Takenori12,Watanabe14}. The ground state properties
(total binding energy, spin parity and one neutron separation energy)
and matter radii of the Ne and Mg isotopes are well reproduced
by the AMD calculation. They reported a sudden rise
in the quadrupole deformation, $\beta_{2}$,
as the Nilsson orbitals originating from the spherical
$0f_{7/2}$ shell gets filled for both the Ne and Mg isotopes,
for $N=19$--28. Given that deformation
would change the nuclear density profile at
or near the nuclear surface, the nuclear radius would
also see a correlated increase.
This was confirmed experimentally
  in systematic studies of the total reaction or interaction cross sections
  of Ne and Mg isotopes on a carbon target in
Refs.~\cite{Takechi10,Takechi10b,Takechi13,Takechi14}.  
  We remark that in addition to the AMD approach with the
  Gogny-D1S interaction~\cite{Minomo11,Minomo12,Takenori12,Watanabe14},
  Refs.~\cite{Horiuchi12,Horiuchi15}
  investigated the nuclear radii by using Skyrme-type effective interactions
  and showed that large quadrupole deformation is essential to
  describe those nuclei.

The extraction of the nuclear density profile is indeed a challenging issue.
Traditionally, electron scattering has been used to measure
the proton density profile~\cite{deVries87}
but it is difficult to extract the neutron density
distribution even for stable nuclei~\cite{Abrahamyan12}.
In this context, proton-nucleus scattering
has been applied successfully~\cite{Sakaguchi17}
to deduce the matter density distribution.
Ref.~\cite{Hatakeyama18} discussed the high-energy
nucleon-nucleus scattering as an effective tool to analyze the nuclear surface
diffuseness. They showed that the information about the half-radius of
the nuclear density profile is encoded in the first diffraction peak
of the nucleon-nucleus elastic scattering differential cross section.
Another extension of proton-nucleus scattering is to deduce
the information about matter density distribution of unstable nuclei
using inverse kinematics~\cite{Matsuda13}. Therefore, this motivates
us to investigate the relationship between the nuclear surface diffuseness
and the spectroscopic information of nuclei,
in the island of inversion, utilizing high-energy nucleon-nucleus scattering. 

In this paper, we deduce the nuclear surface diffuseness
of Ne and Mg isotopes in a systematic way.
For this purpose, we use the two-parameter Fermi density (2pF) distribution,
which defines the nuclear diffuseness~\cite{Hatakeyama18}.
The radius and diffuseness parameters in the 2pF distribution
for neutron-rich Ne and Mg isotopes
are determined so as to reproduce a realistic density distribution
calculated with the antisymmetrized molecular dynamics (AMD).
We then perform a systematic analyses
to find the correlation between the nuclear surface diffuseness
and various structure information of the neutron-rich Ne and Mg isotopes.
Feasibility of extracting the diffuseness parameter using the proton-nucleus
elastic scattering~\cite{Hatakeyama18}
is verified for this mass region by using the Glauber model.

This paper is organized in the following way.
In Section~\ref{formalism.sec},
we give the requisite details of the AMD model relevant for our calculations.
We also briefly explain the formalism of nucleon-nucleus collision
at high incident energy within the Glauber model,
wherein the elastic scattering differential cross sections are evaluated.
The results and discussions on the neutron-rich Ne and Mg isotopes
appear in Sec.~\ref{Results.sec},
followed by the conclusions in Sec.~\ref{Conclusions.sec}.

\section{Theoretical formalism}
\label{formalism.sec}

\subsection{Density and occupation numbers
  from antisymmetrized molecular dynamics}
\label{amd.sec}

We use AMD as a nuclear structure model to calculate the density and occupation numbers of Ne and Mg
isotopes. We start with an $A$-body Hamiltonian
\begin{align}
 H=\sum_{i=1}^At_i - t_{\mathrm{cm}} + \sum_{i<j}^A v_{ij},
\end{align}
where $v_{ij}$ denotes the Gogny D1S density functional plus Coulomb interaction. The center-of-mass
kinetic energy $t_{\mathrm{cm}}$ is subtracted without approximation. 

The variational wave function is the parity-projected Slater determinant of nucleon wave packets
\begin{align}
 \Phi^\pi = P^\pi\mathcal{A}\set{\varphi_1\cdots\varphi_A},
\end{align}
where $P^\pi$ denotes the parity ($\pi=\pm$) projector. The nucleon wave packets has a Gaussian form
\begin{align}
 \varphi_i =& \prod_{\sigma=x,y,z}\exp\set{-\nu_\sigma(r_\sigma - Z_{i\sigma})^2}\nn\\
 &\times\left(a_i\chi_{\frac{1}{2},\frac{1}{2}} + b_i\chi_{\frac{1}{2},-\frac{1}{2}}\right)(\ket{p} \mathrm{or} \ket{n}).
\end{align}
The centroids $\bm Z_i$, width $\bm \nu$ and the spin direction $a_i$ and $b_i$ of the wave
packets are variational parameters. They are determined by minimizing
the following energy with $\beta$ constraint term
\begin{align}
  E(\beta) = \frac{\braket{\Phi^\pi|H|\Phi^\pi}}{\braket{\Phi^\pi|\Phi^\pi}}
  + v_\beta(\braket{\beta} - \beta)^2,
\end{align}
where the strength of the constraint $v_\beta$
is chosen as sufficiently large value to obtain the
optimized wave function $\Phi^\pi(\beta)$,
which has the minimum energy for each given value of
the deformation parameter $\beta$. 

The optimized wave functions are projected to the eigenstate of the angular momentum and superposed
to describe the ground state
\begin{align}
 \Psi^{J\pi}_{M}=\sum_{iK}g_{iK}P^J_{MK}\Phi^\pi(\beta_i),
\end{align}
where the deformation parameter $\beta$ is employed as a generator coordinate. The coefficients
$g_{iK}$ and the ground state energy are obtained by solving the Hill-Wheeler
equation~\cite{Hill53}.

The point nucleon densities are calculated from the ground state wave functions as
\begin{align}
 \rho_{JM}(\bm r) &= 
 \braket{\Psi^{J\pi}_{M}|\sum_i\delta^3(\bm r_i - \bm r_{cm} -\bm r )|\Psi^{J\pi}_{M}} \nn\\
 &=\sum_{l}C^{JM}_{JMl0}\rho^{l}_{J}(r)Y_{l0}(\hat r),
\end{align}
where $\bm r_{cm}$ denotes the center-of-mass coordinate. The $l=0$ component of the density 
$\rho^l_J(r)$ has been used as an input for the Glauber calculation, although the odd-mass nuclei
can have the non-spherical densities with $l\neq 0$. 

The occupation numbers of the $sd$ and $pf$-shells are evaluated in the same manner with
Ref.~\cite{Vishal20, Yoshiki21}. 
First, we choose the single Slater determinant $\Phi^\pi(\beta)$ which has the
maximum overlap with the ground state wave function
$|\braket{P^J_{MK}\Phi^\pi(\beta)|\Psi^{J\pi}_M}|^2$ , and regard it as an approximate 
ground state.  This approximation may be reasonable as the maximum value of the overlap were larger
than 0.90 for all nuclei. Then, we calculate the neutron single-particle wave functions
$\widetilde{\varphi}_i$ of the  approximate ground state wave function, and consider the multipole
decomposition
\begin{align}
 \widetilde{\varphi_i}(\bm r) = \sum_{jlm}\widetilde{\varphi}_{i;jlm}(r) 
 \left[Y_l(\hat r)\times\chi_{\frac{1}{2}}\right]_{jm}.
\end{align}
The norm of $\widetilde{\varphi}_{i;jlm}(r)$ gives us an estimate of the neutron occupation
numbers. The number of the neutron particles in $pf$ orbit is given as
\begin{align}
 m(p),\ m(f) = \sum_{ijm}\braket{\widetilde{\varphi}_{i;jlm}|\widetilde{\varphi}_{i;jlm}} 
 - n_{\mathrm{core}}, 
\end{align}
where $n_\mathrm{core}$ is taken as $n_{\mathrm{core}}=6$ for the $1p$ $(l=1)$ orbit
and 0 for the $0f$ $(l=3)$ orbit
as we assume the complete filling of the $0p$ orbits
by the inert core. In the same
manner, the number of holes in the $sd$ orbits relative to the $N=20$ shell closure is given as 
\begin{align}
 n(sd) = n_{\mathrm{core}} -
 \sum_{l=0,2}\sum_{ijm}\braket{\widetilde{\varphi}_{i;jlm}|\widetilde{\varphi}_{i;jlm}},
\end{align}
where $n_{\mathrm{core}}=14$ for the assumption of the complete filling
of the $0s_{1/2}$ and $0d_{5/2}$ orbits.

\subsection{Nucleon-nucleus reactions with Glauber model}
\label{reaction.sec}

A powerful description of high-energy nuclear reactions
was introduced by Glauber~\cite{Glauber}.
In the collision of a nucleon-nucleus system
within the eikonal and adiabatic approximations,
the elastic scattering amplitude of the Glauber model
including the nuclear ($e^{i\chi_N}$)
and the elastic Coulomb ($e^{i\chi_{C}}$) phase-shift functions
can be calculated by \cite{Suzuki03} 	
	\begin{equation}
	F(\bm{q}) = \dfrac{iK}{2 \pi}\int d\bm{b}\,e^{-i\bm{q}\cdot\bm{b}}
	(1-e^{i\chi_{N}(\bm{b})+i\chi_{C}(\bm{b})}),
	\end{equation}	
	where $\bm{q}$ is the momentum transfer vector, $K$ is incident relativistic wave number 
	corresponding to the projectile-target relative motion, and $\bm{b}$ is the impact parameter vector. 
	The elastic scattering amplitude can be further simplified as
	\begin{eqnarray}    
	F(\bm{q}) = && ~ e^{2i\eta \ln(2Kr_c)}\Bigg[ F_{c}(\bm{q})+ \nonumber\\ 
	&&\dfrac{iK}{2 \pi}\int d\bm{b}\,e^{-i\bm{q}\cdot\bm{b}+2i\eta \ln(\bm{b})}    
	(1-e^{i\chi_{N}(\bm{b})})\Bigg],
	\end{eqnarray}
        where $r_c$ is the distance beyond which the Coulomb potential
        is switched off, whereas incidentally the differential cross section
        does not depend on $r_c$, the Rutherford scattering amplitude 
	\begin{equation}
	F_{c}(\bm{q}) = -\dfrac{2K\eta}{\bm{q}}e^{-2i\eta \ln(\sin(\theta/2))+2i\sigma_{0}},
	\end{equation}
	with $\theta$ as the center of mass scattering angle, $\eta$ as the Sommerfeld parameter, and  $\sigma_{0}$ = arg$\Gamma(1+i\eta)$.
 The elastic scattering differential cross section can then be calculated using 
\begin{equation}
\frac{d\sigma}{d\Omega} = |F(\bm{q})|^{2}.
\end{equation}	
In general, the evaluation of the nuclear phase-shift function 
is demanding because it contains multiple integrations. However, we employ the optical-limit approximation (OLA)~\cite{Glauber, Suzuki03}
for the sake of simplicity.
In the OLA, the multiple scattering effects are ignored by
taking only the leading order term of the cumulant expansion
of the original phase-shift function.
In the case of proton-nucleus scattering, the OLA works well, as demonstrated
in Refs.~\cite{Varga02,Ibrahim09,Hatakeyama14,Hatakeyama15,Nagahisa18,Hatakeyama18}.
The optical phase-shift function for the nucleon-nucleus scattering
in the OLA is given by 	
\begin{equation}
e^{i\chi_N(\bm{b})} \approx \exp\left[
-\int d\bm{r} \rho_{N}(\bm{r}) \Gamma_{NN}(\bm{b}-\bm{s})\right],
\end{equation}
where $\bm{r} = (\bm{s},z)$,  and $\bm{s}$ is
the two-dimensional vector orthogonal to the incident beam direction $z$.
$\rho_{N}(\bm{r})$ denotes the density distributions
of the target nucleus. The profile function  $\Gamma_{NN}$
for the nucleon-nucleon scattering
is incident energy dependent and 
is usually parameterized as given in Ref.~\cite{Lray}:	
\begin{eqnarray}
\Gamma_{NN}(\bm{b})=\dfrac{1-i\alpha_{NN}}{4 \pi \beta_{NN}}
\sigma_{NN}^{\rm tot}\exp\left(-\dfrac{\bm{b}^2}{2 \beta_{NN}}\right),
\end{eqnarray}	
where $\alpha_{NN}$ is the ratio of the real part
to the imaginary part of the nucleon-nucleon scattering amplitude
in the forward direction,
$\beta_{NN}$ is the slope parameter of the differential cross section,
and $\sigma_{NN}^{\rm tot}$ is the nucleon-nucleon total cross section.
Standard parameter sets of the profile function
are listed in Refs.~\cite{Horiuchi07,Ibrahim08}.

\subsection{Nuclear diffuseness}

Let us now define the nuclear surface diffuseness used in this paper.
We assume that the nuclear matter density profile with the mass number $A$
being approximated by a two-parameter Fermi (2pF) distribution as	
\begin{equation}
\rho_{2pF}(r) = \dfrac{\rho_{0}}{1+\exp\left[(r-R)/a\right]}, 	
\end{equation}	  
where $R$ and $a$ are the radius and diffuseness parameters, respectively.
The $\rho_{0}$ value is uniquely determined for a given $R$ and $a$ by
the normalization condition, $\int \rho(r)d\bm{r} = A$.
  Note that nuclear deformation
  induces in general more diffused nuclear surface compared
  to a spherical one~\cite{Horiuchi21}
  and most of the Ne and Mg isotopes considered here
  are deformed~\cite{Takenori12, Horiuchi12, Watanabe14}.
As prescribed in Ref.~\cite{Hatakeyama18},
the nuclear surface density profile,
even though they are deformed,
can be described fairly well by taking the $R$ and $a$ values
so as to reproduce the first peak position and its magnitude of the
elastic scattering differential cross section.
Later we will verify that approach for the application to the
neutron-rich Ne and Mg isotopes.

\begin{figure}[h]
 \begin{center}    
  \includegraphics[width=\linewidth]{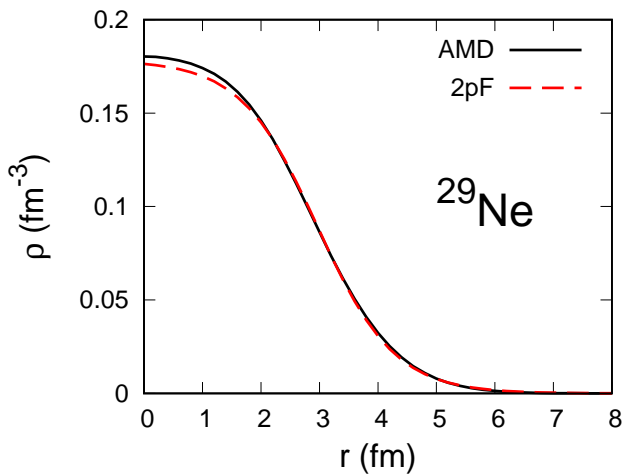}
  \caption{Nuclear matter density distributions of $^{29}$Ne by AMD and the one approximated by the
  2pF function. } \label{density_Ne.fig} 
 \end{center}
\end{figure}

Meanwhile, we evaluate the diffuseness parameters directly
from  any structure model densities by minimizing the quantity as
\begin{equation}
\dfrac{4\pi}{A}\int_0^{\infty} |\rho(r)-\rho_{\rm 2pF}(r)|r^2 dr,
\label{minimize.eq}
\end{equation}
where $\rho$ is the point matter density distribution obtained
with a structure model calculation.
Figure~\ref{density_Ne.fig} shows an example of 
the point matter density distribution of $^{29}$Ne
obtained with AMD, which exhibits
large quadrupole deformation $\beta_2=0.445$~\cite{Takenori12}.
Though the 2pF distribution deviates in the internal region
at $r \lesssim 2$ fm,
it nicely describes the AMD density distributions
around the nuclear surface from $\approx 2$--4 fm.
Hereafter we use the diffuseness parameters obtained directly from
the AMD densities unless otherwise mentioned.

\section{Results and Discussions}
\label{Results.sec}

\subsection{Evolution of the nuclear surface diffuseness for Ne and Mg isotopes}
\label{2pF_mini.sec}
\begin{figure}[h]
 \begin{center}    
  \includegraphics[width=\linewidth]{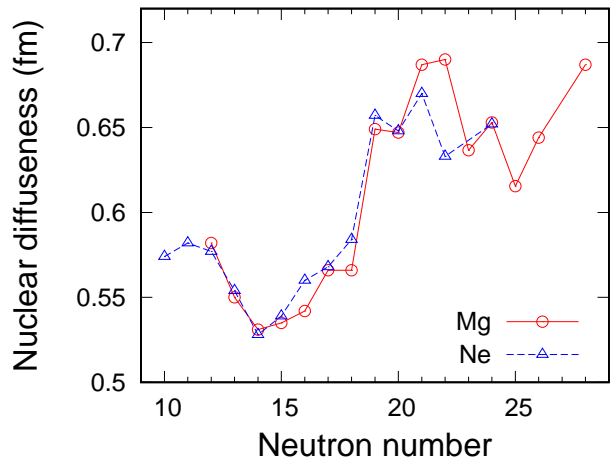}
  \caption{Nuclear surface diffuseness of Ne and Mg isotopes as a function of neutron number.}
  \label{diff_NeMg.fig}
 \end{center}
\end{figure}

\begin{table*}[ht]
  \caption{Occupation numbers of Ne and Mg isotopes with $N\geq 18$. See text for details.}
  \label{occ.tab}
 \begin{ruledtabular}
  \begin{tabular}{cccccccccccc}
       &&& \multicolumn{4}{c}{Ne}&&\multicolumn{4}{c}{Mg}\\
\cline{4-7}\cline{9-12}
$N$ & $J^\pi$ &&  $m(p)$ &$m(f)$&$m(pf)$&$n(sd)$
&&   $m(p)$ &$m(f)$&$m(pf)$&$n(sd)$\\
\hline
18 &$0^+$  &&$-$0.03&0.08&0.05&2.12&&0.24&0.27&0.50&2.66\\
19 &$1/2^+$&& 0.82&1.16  &1.98&3.26&&0.76&1.18&1.94&3.36\\
20 &$0^+$  && 0.84&1.14  &1.97&2.26&&0.76&1.21&1.97&2.27\\
21 &$3/2^-$&& 0.97&1.93  &2.91&2.28&&0.92&1.95&2.87&2.39\\
22 &$0^+$  && 0.97&1.72  &2.69&1.07&&1.12&2.70&3.81&2.39\\
23 &$3/2^+$&&  -- &-- &--  &--  &&1.01&2.86&3.87&1.26\\
24 &$0^+$  &&1.02&2.90   &3.91&0.22&&0.98&2.94&3.92&0.24\\
25 &$5/2^-$&&--  &--     &--  &--&&1.07&3.82&4.89&0.24\\
26&$0^+$  &&--  &--      &--  &--&&1.40&4.46&5.87&0.30\\
28 &$0^+$  &&--  &--     &--  &--&&1.97&5.78&7.75&0.35\\
  \end{tabular}
 \end{ruledtabular}
\end{table*}

\begin{figure}[ht]
 \includegraphics[width=\hsize]{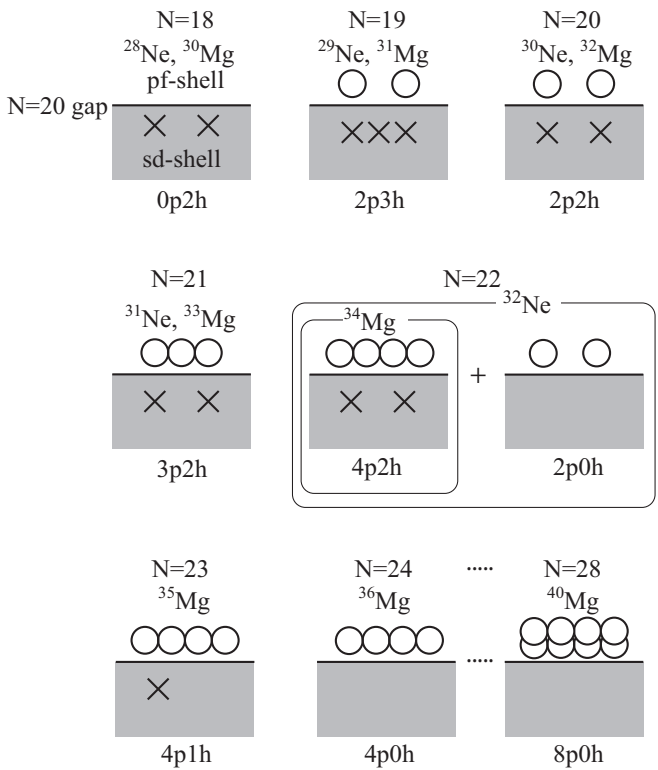}
 \caption{Schematic illustrations of the $m$p$n$h configurations relative to the $N=20$ shell
 closure. The circles indicate the particles in $pf$ shell while the crosses indicate the holes in
 $sd$ shell.} 
 \label{mpnh.fig}
\end{figure}

\begin{figure}[ht]
 \includegraphics[width=0.8\hsize]{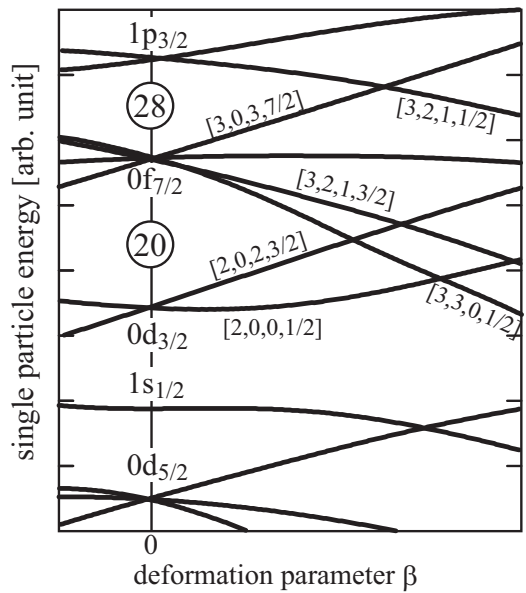}
 \caption{A schematic Nilsson diagram for prolate deformation.} 
 \label{nilsson.fig}
\end{figure}

Figure~\ref{diff_NeMg.fig} plots the surface diffuseness of Ne and Mg isotopes extracted from the
densities obtained by AMD. Reflecting the similarity in the nuclear
deformations~\cite{Takenori12,Watanabe14},
the diffuseness parameter of Ne and Mg isotopes also
show similar dependence on neutron number.
We found that the neutron occupation of the weakly bound
$1p_{3/2}$ orbit strongly influences the global behavior of the diffuseness parameter. To elucidate
this point, Table~\ref{occ.tab} lists the number of neutrons in the 
$pf$ orbits and holes in the $sd$ orbits for nuclei with $N\geq 18$,
and  Figure~\ref{mpnh.fig} illustrates the dominant particle-hole configuration of each nucleus
estimated from Table~\ref{occ.tab}.

The nuclei up to $N=18$ have approximately zero particles in $pf$ orbits, which is the normal filling
expected from the ordinary shell structure. Consequently, their diffuseness parameters are 
close to the standard value of 0.54 fm~\cite{BM}.  The particle-hole configuration is drastically
changed in the island of inversion because of the loss of the magic number $N=20$. 
The ground states of $N=19$ nuclei, 
$^{29}{\rm Ne}$ and $^{31}{\rm Mg}$, are dominated by a 2p3h configuration in which two neutrons
are promoted into the $pf$ orbits across the $N=20$ shell gap. This intruder configuration induces
strong quadrupole deformation and the mixing of the $f$- and $p$-waves. Consequently, these nuclei
have sizable occupation numbers of the $1p_{3/2}$ orbit
(0.82 in $^{29}{\rm Ne}$ and 0.76 in
$^{31}{\rm Mg}$) as well as $0f_{7/2}$. Note that the $1p_{3/2}$
orbit is located above the $N=28$ shell gap
in stable nuclei. Therefore, the occupation of the $1p_{3/2}$ orbit means that the magic numbers 20
and 28 are simultaneously lost in the island of inversion. Since the $1p_{3/2}$ orbit has large
diffuseness, the density 
distributions of the $N=19$ nuclei are also diffused compared to $N=18$ nuclei.  This situation may
be more simply explained by the Nilsson orbits illustrated in   Fig.~\ref{nilsson.fig}. In the $N=19$
nuclei, the two neutrons occupy an intruder orbit  with the asymptotic quantum number
$[N,N_z,\Lambda,\Omega]=[3,3,0,1/2]$, which originates in the spherical $0f_{7/2}$ orbit. Because of
the deformation and weak binding, this orbit is an admixture of the $p$- and $f$-waves. So, the
intruder orbit $[3,3,0,1/2]$ is the cause of large diffuseness of $N=19$ nuclei.   

The isotopes from $N=19$ to $N=21$ ($^{31}{\rm Ne}$) and $N=22$ ($^{33}{\rm Mg}$) are also
regarded as the nuclei in the island of inversion because their ground states are also dominated by
the intruder $m$p$n$h configurations with $m,n >0$. Similarly to the $N=19$ nuclei,  strong
deformation mixes the $f$- and $p$-waves and increases the diffuseness. In terms of the Nilsson
orbit, the intruder $[3,3,0,1/2]$ and $[3,2,1,3/2]$ orbits are playing a role for diminishing the
$N=20$ and 28 shell gaps and creating the island of inversion.   

At $^{35}{\rm Mg}$ $(N=23)$, the intruder orbits $[3,3,0,1/2]$ and $[3,2,1,3/2]$ are fully
occupied, and the holes in the $sd$ orbits start to be filled. Because the $sd$ orbits are deeply
bound, they partially cancel out the diffuseness increased by $1p_{3/2}$. As a result, the
diffuseness parameter slightly reduces toward $^{37}{\rm Mg}$ $(N=25)$.   At $N=24$, the holes in
$sd$-shell are completely filled, and hence one may regard $^{36}{\rm Mg}$ as the border of the
island of inversion. 

In the $N=26$ and 28 nuclei, another intruder orbit $[3,2,1,1/2]$ which originates in the spherical
$1p_{3/2}$ orbit comes down and is inverted with the orbit $[3,0,3,7/2]$ leading to the explicit
loss of the $N=28$ magicity~\cite{Yoshiki21,Hamamoto09,Hamamoto16}. 
Since this intruder orbit is also an admixture of the $p$- and
$f$-waves, the occupation number of $1p_{3/2}$ gradually increases resulting in the growth of the
diffuseness toward $^{40}{\rm Mg}$.  Thus, the global behavior of the diffuseness parameter can be
explained by the occupation of $1p_{3/2}$. 

Let us now comment on the structure and diffuseness parameters of these isotopes. Firstly, we note
that $N=22$ nuclei, $^{34}{\rm Mg}$ and $^{32}{\rm Ne}$ have slightly different diffuseness
parameters in the present calculation. We found that  $^{34}{\rm Mg}$ is dominated by a 4p2h configuration, while $^{32}{\rm Ne}$ is an admixture of a 4p2h and a 2p0h configurations. Hence, 
$^{34}{\rm Mg}$ has larger diffuseness parameter than $^{32}{\rm Ne}$. Secondly, we note that the
spin-parity of $^{29}{\rm Ne}$, $^{35}{\rm Mg}$  and $^{37}{\rm Mg}$ have not been firmly determined
yet. For example, our calculation suggests the $1/2^+$ ground state of $^{29}{\rm Ne}$, 
while a shell model calculation suggests the $3/2^+$ ground
state~\cite{Tripathi05,Tripathi06}. Contrary to these theoretical results, the $3/2^-$ ground state
was suggested by a Coulomb breakup experiment~\cite{Kobayashi14}. As different spin-parity means
different particle-hole configurations, we expect that more detailed analysis of the diffuseness
will identify the spin-parity of these nuclei. However, to make our discussion transparent, we only
adopted the spin-parity calculated by AMD. 

\subsection{Single-particle model analysis for nuclear diffuseness}

Based on the spectroscopic information obtained from the AMD wave function,
here we attempt to understand the large nuclear diffuseness values for $N>18$
through a single-particle model approach.
Assuming that an $N=18$ isotope is a core nucleus
with 0p2h configuration,
we consider multi-particle-multi-hole ($\bar{m}$p$\bar{n}$h) configurations
for the valence neutrons according to the dominant configurations
illustrated in Fig.~\ref{mpnh.fig}.
The normalized valence neutron orbits $\phi(nl_j)$
are generated by the following
core-neutron potential~\cite{BM}
\begin{align}
  U=V_0f(r)+V_1r_0^2 \bm{l}\cdot\bm{s}\frac{1}{r}\frac{d}{dr}f(r).
\end{align}
The Woods-Saxon form factor $f(r)=\{1+\exp[(r-R_c)/a_c]\}^{-1}$
is employed. We take $R_c=r_0A_c^{1/3}$, where
$r_0=1.25$ fm, $A_c=28$ (30) for the Ne (Mg) isotopes with $N\geq 19$,
and  $a_c=0.75$ fm. The spin-orbit strength is taken to follow
the systematics~\cite{BM} $V_1=18.0$ (19.2) MeV for the Ne (Mg) isotopes.
This parameter set reasonably reproduced
the level structure at around the island of inversion~\cite{Horiuchi10}.
We generate the single-particle wave functions
by varying $V_0$ and construct the nuclear density as
\begin{align}
  \rho&=\rho_c(N_c=18)+\rho_v(N_v),
\end{align}
where $\rho_c (N_c=18)$ is the density distribution of $^{28}$Ne
or $^{30}$Mg obtained by the AMD calculation,
and $\rho_v(N_v)$ is the density distribution of the valence neutrons
defined by
\begin{align}
\rho_v&=\bar{m}\left[\alpha|\phi(1p_{3/2})|^2+
  (1-\alpha)|\phi(0f_{7/2})|^2)\right]\notag\\
&-\bar{n}|\phi(0d_{3/2})|^2.
\label{rhov.eq}
\end{align}
In this model, the number of the valence neutrons
satisfies $N_v=\bar{m}-\bar{n}$.
The $\bar{m}$ is the number of particle states
that shares the $1p_{3/2}$ and $0f_{7/2}$
orbits with mixing probability $\alpha=m(p)/m(pf)$ listed in Tab~\ref{occ.tab}.
The $|\bar{n}|$ describes the number of the hole ($\bar{n}>0$)
or particle state ($\bar{n}<0$)
measured from the core nucleus ($N=18$). The particle
or hole state in the $sd$ shell is assumed to be $0d_{3/2}$.
Here we take the most plausible configuration for each isotope,
which corresponds to the $\bar{m}$p($\bar{n}-2$)h
configurations drawn in Fig.~\ref{mpnh.fig} according
to the spectroscopic information of the AMD wave function.
More specifically,
we take $(\bar{m},-\bar{n})=(2,-1)$, (2,0), (3,0), (3,1), and (4,2)
for $^{29-32,34}$Ne, and
$(\bar{m},-\bar{n})=(2,-1)$, (2,0), (3,0), (4,0), (4,1), (4,2),
(5,2), (6,2), and (8,2) for $^{31-38,40}$Mg, respectively.
Finally, the potential strength $V_0$
is fixed for each isotope so as to reproduce
the root-mean-square (rms)
matter radius obtained by the AMD (They are tabulated
in Tab.~\ref{diff.tab}). This is reasonable because
the behavior of the single-particle wave function
near the nuclear surface crucially depends on its binding energy
and will reflect in the nuclear radius.
This effect can be incorporated in this model
through the adjustment of $V_0$.

 \begin{figure}[h]
	\begin{center}    
	 \includegraphics[width=\linewidth]{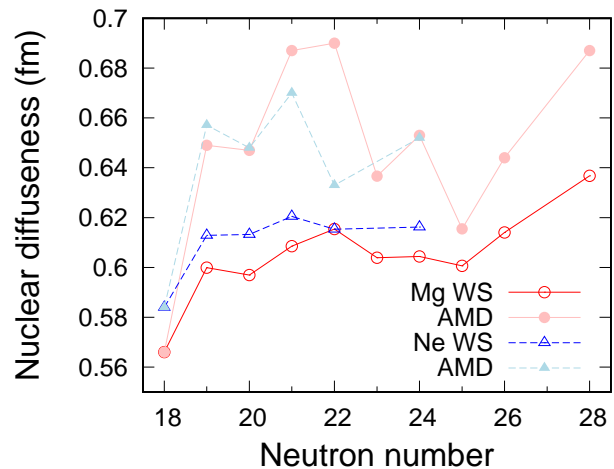}
		\caption{Nuclear diffuseness parameters
                  extracted from density distributions obtained by
                  the single-particle model for $N> 18$. The AMD results
                  (Fig.~\ref{diff_NeMg.fig}) are also
                  plotted for comparison.
            } 
		\label{diffWS.fig}
	\end{center}
\end{figure}

 \begin{figure}[h]
	\begin{center}    
	 \includegraphics[width=\linewidth]{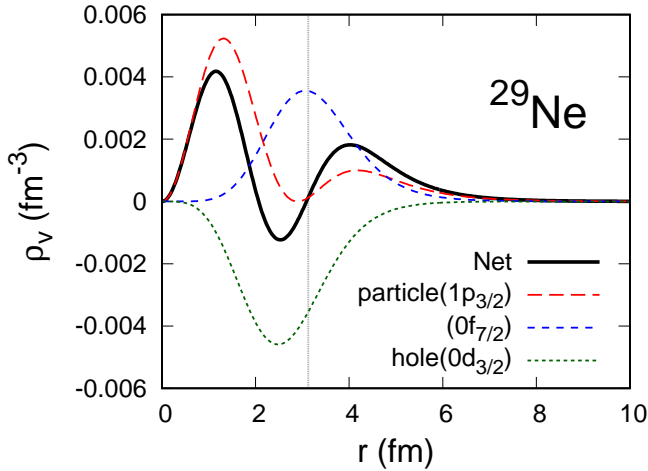}
		\caption{Valence neutron density distribution
                  and its decomposition into particle $1p_{3/2}$ and $0f_{7/2}$
                  states and $0d_{3/2}$ hole state. See text for details.
                  A vertical thin dotted line denotes the radius parameter
                  value 3.13, extracted from the density distribution
                  obtained by the single-particle model. 
                  }
		\label{densWS.fig}
	\end{center}
\end{figure}

 Using those calculated density distributions of the Ne and Mg isotopes
 for $N\geq 19$, we extract the diffuseness parameters
 directly from the model density by using Eq.~(\ref{minimize.eq}).
 Figure~\ref{diffWS.fig} draws these extracted diffuseness parameters
 for $N\geq 19$.  We find overall underestimation of the absolute value.
 This is partly because the AMD wave function is expressed by Gaussian
 wave packets and thus the density distribution at nuclear surface
 changes more sensitive to the occupation numbers.
 On the other hand, the isotope dependence is fairly well described:
 a sudden increase from $N=18$ to 19 and showing a zigzag pattern
 for $N$ increases further.
 Basically, the nuclear diffuseness grows as an increase of
 the number of the particle in the $pf$ shell
 and the hole of the $sd$ shell.
 We will now discuss this feature in detail.
 
 Figure~\ref{densWS.fig} plots the density distributions 
 of the valence neutrons $\rho_v$ of $^{29}$Ne
 obtained by the single-particle model analysis.
 We also plot the decompositions of the valence neutron density
 into the particle $1p_{3/2}$, $0f_{7/2}$ and hole $0d_{3/2}$
 components, which correspond to the first to third terms
 in Eq.~(\ref{rhov.eq}), respectively.
 The diffuseness parameter describes a slope around the
 radius parameter~\cite{Kohama16}
 and is also plotted as a vertical line at 3.13 fm.
 As expected, the single-particle density 
 with the $1p_{3/2}$ orbit exhibits the most extended
distribution just beyond the nuclear radius.
The peak positions of the $0f_{7/2}$ particle and $0d_{3/2}$ hole states
are different. The $0f_{7/2}$ particle state
contributes the increase of the density around the nuclear radius,
while the $0d_{3/2}$ hole plays to reduce the density
below the nuclear radius. Summing up all those contributions, as a net,
the valence neutron density reduces (increases) the density below (beyond)
the nuclear radius, leading to
the large diffuseness of the nuclear surface at $N=19$
compared to $N=18$.
This is consistent with the finding of
  Ref.~\cite{Horiuchi21b}, in which contributions of the single-particle
orbits to the nuclear surface diffuseness were discussed in detail.
We also tried to make the same analysis by assuming
the particle $0d_{3/2}$ configuration for the valence neutron
following the spherical shell model filling.
No bound $0d_{3/2}$ orbit was obtained to satisfy the condition of
this single-particle model, and thus this assumption
appears to be unrealistic.

At $N=20$, the diffuseness parameter is reduced from $N=19$
because the hole $0d_{3/2}$ state is filled by addition of a neutron.
It again increases at $N=21$ due to the occupation
of a neutron in the $pf$ shell.
At $N=22$, where the configurations of $^{32}$Ne and $^{34}$Mg
are different. The diffuseness of $^{32}$Ne decreases compared to $^{31}$Ne
due to the occupation of the particle $0d_{3/2}$ state, whereas
for $^{34}$Mg it increases due to the $pf$ shell filling.
For $N>22$, since the $sd$ shell is fully occupied,
the nuclear diffuseness gradually increases towards $N=28$,
showing a small kink at $N=25$.

Though the present single-particle model analysis
  is somewhat qualitative, the evolution of the nuclear diffuseness
tells us a variety of the structure information.
A systematic determination of
the nuclear diffuseness is interesting
as it includes the spectroscopic information,
which is essential in describing the exotic nuclear states
in the island of inversion. 

\subsection{Extraction of the nuclear diffuseness from the reaction observables}
Here we extract the nuclear diffuseness from the elastic scattering
differential cross sections for the Ne and Mg isotopes.
The unknown radius and diffuseness parameters
are evaluated by using the elastic scattering differential
cross section of a nucleus-proton reaction calculated with
the Glauber model following the prescription given in Ref.~\cite{Hatakeyama18}.
First we compute the elastic scattering differential cross sections using
realistic density distributions calculated with AMD.
We then demand the $R$ and $a$ values
of the 2pF distribution reproducing both the first peak position and
its magnitude of the elastic scattering cross sections.

Table~\ref{diff.tab} lists the resulting $a$ values obtained
at various incident energies.
We use a set of the parameters of the profile function
in Ref.~\cite{Horiuchi07} and choose the incident energies
of 325, 550, and 800 MeV, where the isospin dependence of the nucleon-nucleon
cross section is neglected~\cite{Hatakeyama18}.
As was shown in Ref.~\cite{Hatakeyama18}, the extracted
$a$ values do not depend much on the incident energy.
We also did the same analysis
  without the elastic Coulomb contribution.
  The results are listed in parentheses in Table~\ref{diff.tab},
  and indicate that the elastic Coulomb contributions are negligible,
  as expected.
  
To verify this approach for the neutron-rich Ne and Mg isotopes,
we compare the diffuseness parameters obtained directly from the AMD densities (see section \ref{2pF_mini.sec}).
As shown in Table~\ref{diff.tab},
the resulting diffuseness parameters indicated by ``AMD''
also agree with those obtained by the elastic scattering diffraction.
The nuclear diffuseness of the neutron-rich Ne and Mg isotopes can be extracted 
as a robust quantity by measuring the nucleus-nucleon elastic scattering
differential cross sections at the first peak position.

Table~\ref{diff.tab} also lists the rms point matter radii simultaneously
obtained by the analysis of the elastic scattering cross section.
We also see good agreement between the results extracted
from this analysis and the ``AMD'' results.
However, it should be noted that
the rms matter radius of the 2pF distribution tends to overestimate
that of the original AMD density distributions (shown as AMD$^*$ in Table~\ref{diff.tab}) at most by $\approx 0.1$ fm
or typically in $\approx 3$\%.
Since the tail of the AMD density drops as a Gaussian,
the 2pF model density is not very appropriate to describe the tail
regions of the AMD density, which contributes the rms radius.
Actually, in the analysis of the single-particle model using the correct
asymptotic tail, the deviation of the rms matter radii
between the original and 2pF density distributions
is reduced typically in $\approx 1$\%.
In principle, both values can be determined more accurately
by the 2pF distribution obtained from the first peak position
and its magnitude of the elastic scattering
differential scattering cross section
as the rms radius and diffuseness are nicely
reproduced when the density distributions of the Hartree-Fock
method on grid points within $\approx 1$\%~\cite{Hatakeyama18}.
In a practical experimental situation,
where only the peak position and its magnitude of
the elastic scattering cross section is known,
the matter radius and diffuseness can be determined by the 2pF
model density within a certain accuracy.

\begin{table*}[ht]
\centering
\caption{Nuclear surface diffuseness, $a$,
  and rms point matter radii, $r_m$, of Ne and Mg isotopes
  extracted from the elastic scattering differential cross sections
  at various incident energies $E$ in MeV. The values in parentheses
    are the results without the elastic Coulomb phase. 
    The $a$ values extracted directly by minimizing the difference between the AMD and 2pF density distributions (see section \Ref{2pF_mini.sec}) are listed as AMD for comparison.
    AMD$^*$ shows the rms point matter radii calculated with the original AMD density distributions.   
}
 	\label{diff.tab} 	
 		\begin{tabular}{ccccccccccc}
 		  \hline\hline
 		  &\multicolumn{4}{c}{$a$ (fm)}&& \multicolumn{5}{c}{$r_m$ (fm)}\\
\cline{2-5}\cline{7-11}                 
nucleus  &  $E=325$ &  550 &  800  &  AMD  && $E=325$ &  550 &  800  &  AMD &AMD$^*$ \\
\hline 			
$^{20}$Ne &  0.579(0.582) & 0.578(0.579) & 0.571(0.570) & 0.574&&3.049(3.044)&3.039(3.040)&3.031(3.025)&2.984&2.925\\
$^{21}$Ne &  0.586(0.590) & 0.585(0.586) & 0.577(0.577) & 0.582&&3.088(3.089)&3.078(3.079)&3.066(3.066)&3.023&2.962\\
$^{22}$Ne &  0.579(0.582) & 0.577(0.578) & 0.570(0.570) & 0.577&&3.104(3.099)&3.088(3.090)&3.082(3.081)&3.043&2.983\\  
$^{23}$Ne &  0.552(0.554) & 0.551(0.551) & 0.544(0.544) & 0.554&&3.076(3.071)&3.071(3.067)&3.060(3.059)&3.030&2.977\\ 
$^{24}$Ne &  0.520(0.522) & 0.519(0.519) & 0.514(0.514) & 0.528&&3.041(3.042)&3.039(3.036)&3.035(3.034)&3.013&2.967\\
$^{25}$Ne &  0.533(0.536) & 0.531(0.532) & 0.524(0.524) & 0.539&&3.094(3.091)&3.083(3.085)&3.076(3.076)&3.049&2.996\\
$^{26}$Ne &  0.558(0.563) & 0.557(0.558) & 0.547(0.547) & 0.560&&3.174(3.172)&3.164(3.164)&3.150(3.151)&3.123&3.049\\
$^{27}$Ne &  0.566(0.571) & 0.565(0.566) & 0.555(0.555) & 0.568&&3.235(3.235)&3.226(3.226)&3.212(3.211)&3.184&3.114\\
$^{28}$Ne &  0.582(0.586) & 0.580(0.581) & 0.570(0.570) & 0.584&&3.317(3.310)&3.302(3.302)&3.288(3.288)&3.265&3.191\\
$^{29}$Ne &  0.658(0.663) & 0.658(0.669) & 0.647(0.647) & 0.656&&3.435(3.429)&3.424(3.423)&3.404(3.405)&3.352&3.282\\
$^{30}$Ne &  0.662(0.669) & 0.661(0.663) & 0.648(0.648) & 0.655&&3.490(3.485)&3.470(3.473)&3.449(3.449)&3.410&3.315\\
$^{31}$Ne &  0.684(0.692) & 0.684(0.686) & 0.670(0.670) & 0.676&&3.563(3.559)&3.547(3.549)&3.525(3.529)&3.475&3.374\\  
$^{32}$Ne &  0.633(0.638) & 0.632(0.633) & 0.621(0.621) & 0.633&&3.509(3.504)&3.499(3.498)&3.482(3.482)&3.450&3.370\\
$^{34}$Ne &  0.646(0.652) & 0.645(0.646) & 0.634(0.635) & 0.652&&3.583(3.582)&3.572(3.571)&3.559(3.565)&3.526&3.438\\
\hline                                                                                                             
$^{24}$Mg &  0.581(0.585) & 0.581(0.582) & 0.574(0.574) & 0.582&&3.164(3.161)&3.158(3.158)&3.150(3.149)&3.109&3.049\\
$^{25}$Mg &  0.547(0.550) & 0.546(0.547) & 0.540(0.541) & 0.550&&3.121(3.118)&3.112(3.113)&3.105(3.110)&3.073&3.028\\
$^{26}$Mg &  0.522(0.524) & 0.521(0.522) & 0.515(0.515) & 0.531&&3.099(3.094)&3.093(3.095)&3.083(3.083)&3.072&3.018\\
$^{27}$Mg &  0.530(0.534) & 0.530(0.530) & 0.524(0.524) & 0.535&&3.135(3.138)&3.134(3.129)&3.128(3.127)&3.095&3.051\\
$^{28}$Mg &  0.533(0.537) & 0.532(0.533) & 0.526(0.526) & 0.542&&3.174(3.173)&3.165(3.165)&3.161(3.161)&3.130&3.082\\ 
$^{29}$Mg &  0.565(0.571) & 0.565(0.567) & 0.556(0.556) & 0.566&&3.280(3.273)&3.265(3.270)&3.254(3.253)&3.233&3.166\\
$^{30}$Mg &  0.560(0.565) & 0.559(0.560) & 0.550(0.550) & 0.566&&3.305(3.297)&3.290(3.289)&3.278(3.278)&3.258&3.191\\
$^{31}$Mg &  0.646(0.653) & 0.646(0.648) & 0.637(0.637) & 0.649&&3.452(3.453)&3.439(3.442)&3.429(3.428)&3.387&3.315\\
$^{32}$Mg &  0.647(0.655) & 0.647(0.650) & 0.637(0.637) & 0.647&&3.485(3.482)&3.466(3.474)&3.453(3.455)&3.419&3.336\\
$^{33}$Mg &  0.688(0.697) & 0.689(0.692) & 0.677(0.677) & 0.687&&3.603(3.598)&3.585(3.592)&3.569(3.568)&3.519&3.428\\
$^{34}$Mg &  0.691(0.701) & 0.692(0.695) & 0.679(0.680) & 0.690&&3.639(3.634)&3.620(3.625)&3.603(3.604)&3.550&3.450\\
$^{35}$Mg &  0.636(0.644) & 0.637(0.638) & 0.626(0.627) & 0.635&&3.545(3.547)&3.539(3.536)&3.522(3.528)&3.482&3.408\\
$^{36}$Mg &  0.648(0.654) & 0.648(0.649) & 0.638(0.638) & 0.653&&3.602(3.595)&3.591(3.588)&3.579(3.579)&3.542&3.466\\
$^{37}$Mg &  0.608(0.614) & 0.607(0.608) & 0.600(0.600) & 0.616&&3.557(3.552)&3.544(3.543)&3.540(3.534)&3.510&3.441\\
$^{38}$Mg &  0.636(0.642) & 0.635(0.637) & 0.628(0.628) & 0.644&&3.646(3.639)&3.632(3.636)&3.622(3.622)&3.590&3.511\\
$^{40}$Mg &  0.680(0.688) & 0.679(0.681) & 0.666(0.666) & 0.687&&3.779(3.776)&3.763(3.765)&3.745(3.745)&3.711&3.620\\
\hline\hline
\end{tabular}                
\end{table*}

\section{Conclusions}
\label{Conclusions.sec}

The island of inversion, in the medium mass region of the nuclear chart, is characterized by intruder configurations in the ground state of nuclei. Apart from resulting in large deformations the change in occupation of nucleons in different energy levels will impact the nuclear density profile, and in particular the nuclear surface diffuseness. In this work, we have discussed the relationship between the nuclear diffuseness and the spectroscopic information of nuclei at or close to the island of inversion, specifically for Ne and Mg isotopes with $N = 19$ to $28$. 

We have calculated the structure of Ne and Mg isotopes using the antisymmetrized molecular dynamics (AMD). We then construct a phenomenological two-parameter Fermi (2pF) density distribution, which has adjustable radius and diffuseness parameters. These parameters are then estimated by minimizing the difference in densities obtained by AMD and the 2pF density distribution. In a complimentary approach the radius and diffuseness parameters, of the 2pF density distribution, are also determined so as to reproduce the first peak position and its magnitude of elastic scattering differential cross section obtained with the AMD densities in the Glauber model. The results obtained with these two approaches mostly agree within a limit of 1\% to 3\%.

The AMD results reveal that there is a drastic increase in the occupation number of neutrons in the $pf$ orbit from $N=19$ onwards, compared to $N=18$,  in Ne and Mg isotopes. This intruder configuration induces a strong deformation owing to the mixing of the $p$- and $f$-orbits and signals the onset of the island of inversion in this mass range. We observed that the occupancy of the neutron in weakly bound $1p_{3/2}$ orbit has a significant impact on the overall behavior of the nuclear diffuseness. Given that the $1p_{3/2}$ orbit has a large diffuseness, nuclei with a sizable neutron occupation number in this orbit is also characterized by a large nuclear surface diffuseness. An estimate of the valence neutron density distribution using a single-particle model, with $^{29}$Ne as a test case, also confirms this conclusion. 

The breakdown of the $N=20$ magic number changes the particle-hole configuration notably in the island of inversion. The 2p3h configuration, is dominant in the ground state of $N=19$ nuclei, in which two neutrons occupy the $pf$ orbits above the $N=20$ shell gap. A similar behavior also observed for the loss of $N=28$ ($^{40}$Mg) magic number due to increasing admixture of the $p$- and $f$-intruder orbits resulting in the increase of diffuseness in $^{38-40}$Mg. However, in $^{35-37}$Mg the filling up of the holes in the $sd$-shell for $N=23$ to $25$  results in the relative reduction of the diffuseness.

We have also shown that the information on nuclear diffuseness of neutron-rich Ne and Mg isotopes can be obtained by calculating the first diffraction peak of nucleon-nucleus elastic scattering differential cross section. As expected, at high energies, the Coulomb contribution, in determining the surface diffuseness and matter radii of these isotopes is very small, typically less than 1\%.

Finally, let us remark that a large surface diffuseness in neutron rich Ne and Mg could have consequences in determining the abundance of these nuclei in explosive nucleosynthsis. In fact, properly accounting for the structure of these exotic medium mass isotopes is a prerequisite in subsequently determining the r-process path in neutron star mergers and in the post-collapse phase of a type II or type Ib supernova \cite{Terasawa01,Shubhchintak13}. 
 
\acknowledgments

This work was in part supported by  
JSPS KAKENHI Grants Nos. 18K03635 and 19K03859,
the collaborative research programs 2021,
Information Initiative Center, Hokkaido University and the
Scheme for Promotion of Academic and Research Collaboration
(SPARC/2018-2019/P309/SL), MHRD, India. V.C. also acknowledges MHRD,
India for a doctoral fellowship and a grant from SPARC to visit
the Hokkaido University.

\end{document}